\documentclass{PoS}

\title{Theoretical update on low energy neutrino--nucleus reactions}

\ShortTitle{Theoretical update on low energy neutrino--nucleus reactions}

\author{\speaker{J. Nieves}%
         \thanks{This research was supported by
  DGI and FEDER funds, under contracts FIS2005-00810, FIS2006-03438,
  FPA2007-65748, and the Spanish Consolider-Ingenio 2010 Programme
  CPAN (CSD2007-00042), by Junta de Andaluc\'\i a and Junta de
  Castilla y Le\'on under contracts FQM0225 and SA016A07, and it is
  part of the EU integrated infrastructure initiative Hadron Physics
  Project under contract number RII3-CT-2004-506078.}, M.J. Vicente-Vacas\\
        Instituto de F\'\i sica Corpuscular (IFIC), Centro Mixto
  CSIC-Universidad de Valencia, Institutos de Investigaci\'on de
  Paterna, Aptd. 22085, E-46071 Valencia, Spain\\
}

\author{J.E. Amaro, M. Valverde\\
Departamento de F{\'\i}sica At\'omica,
Molecular y Nuclear, Universidad de Granada, E-18071 Granada, Spain\\
}

\author{E. Hern\'andez\\
Grupo de F\'\i sica Nuclear, Departamento
de F\'\i sica Fundamental e IUFFyM,\\ Facultad de Ciencias, E-37008
Salamanca, Spain.\\
}

\abstract{We review Quasielastic (QE) inclusive and semi--inclusive
  neutrino/antineutrino Charged Current (CC) and Neutral Current (NC)
  induced nuclear reactions at intermediate energies. We pay special
  attention to nuclear corrections besides Pauli blocking: Long and
  Short range nuclear correlations (RPA and SRC) and particle and hole
  Spectral Functions (SF).  We also critically review the use of the
  Plane and Distorted Wave Impulse approximations (PWIA and DWIA) to
  describe inclusive one nucleon knockout reactions off nuclei. In
  this context, we present results from a Monte Carlo cascade method
  to account for the rescattering of the outgoing nucleon. Finally, we
  examine the effects of chiral non-resonant terms in neutrino pion
  production off the nucleon, and present some preliminary results on
  nuclear coherent pion production induced by neutrinos. }

        \FullConference{10th International Workshop on Neutrino
          Factories, Super beams and
          Beta beams\\
          June 30 - July 5 2008\\
          Valencia, Spain}

\begin{document}

\section{Introduction}
Neutrino physics is at the forefront of current theoretical and
experimental research in astro, nuclear, and particle physics. The
presence of neutrinos, being chargeless particles, can only be
inferred by detecting the secondary particles they create when
colliding and interacting with matter. Nuclei are often used as
neutrino detectors, thus the interpretation of neutrino data heavily
relies on detailed and quantitative knowledge of the features of the
neutrino-nucleus interaction. At low and intermediate energies, the
neutrino-nucleus cross section is dominated by QE and single pion
production processes. Those processes are largely dominated by
mechanisms where the gauge boson ($W^{\pm}, Z^0$) inside the nuclear
medium is absorbed by one nucleon, or excites a $\Delta(1232)$
resonance which subsequently decays into a $N\pi$ pair, respectively.
There is a general consensus among the theorists that a simple Fermi
Gas (FG) model, widely used in the analysis of neutrino oscillation
experiments, fails to provide a satisfactory description of the
measured cross sections, and inclusion of further nuclear effects is
needed~\cite{sakuda}. In the first part of the talk, I will focus on
the most relevant nuclear ingredients affecting to QE inclusive and
semi-inclusive processes. Next, I will examine the structure of the
neutrino pion production off the nucleon amplitude, and the role
played by chiral symmetry.

\section{QE Inclusive and Semi-Inclusive Reactions}

\begin{figure}[h]
\begin{center}
 \includegraphics[height=4.cm,width=6cm]{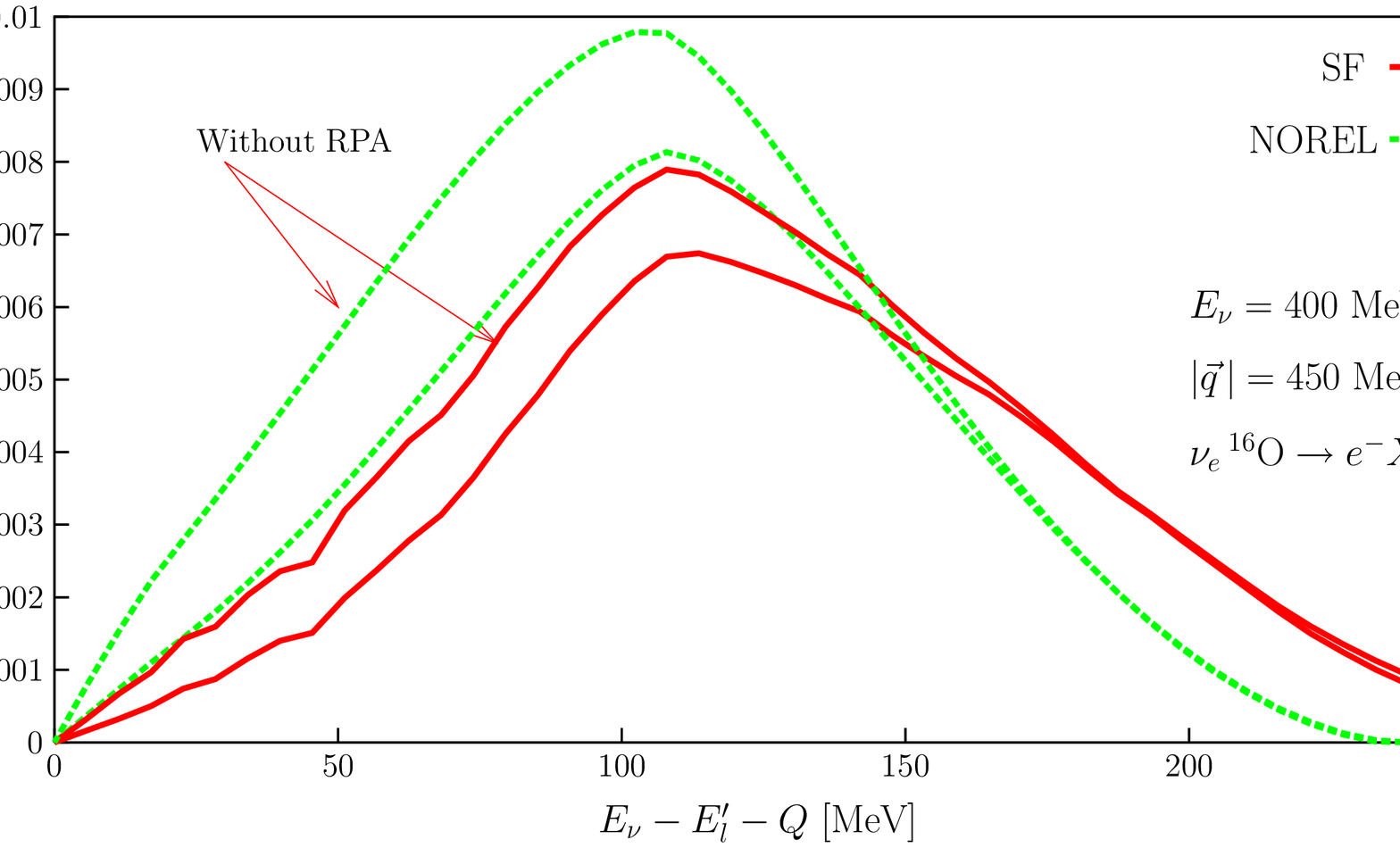}\hspace{2cm}
\includegraphics[height=4.cm,width=6cm]{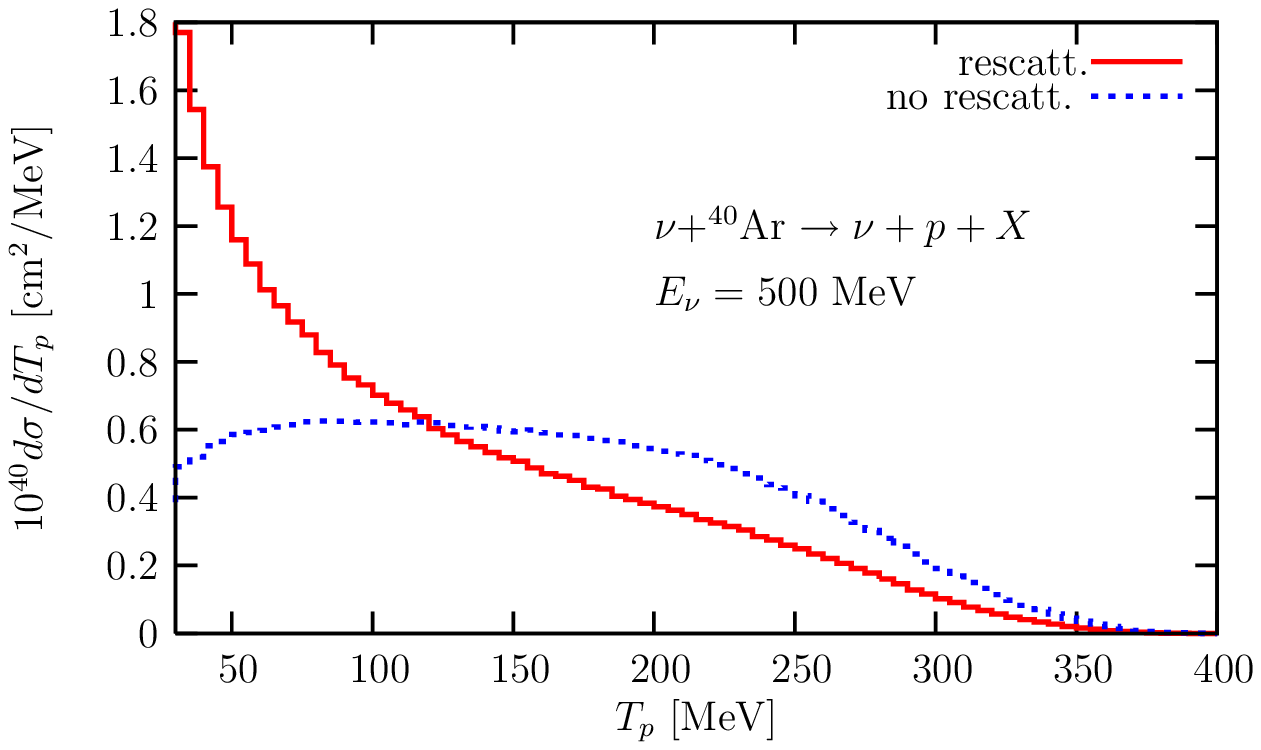}
\end{center}
\caption{\footnotesize Left: $\nu_e$ inclusive QE differential cross
  sections in $^{16}$O as a function of the transferred energy, for a
  fixed transferred momentum. We show results with and without RPA and
  SRC and with (SF) and without (NOREL) SF+FSI effects.  Right:
  $^{40}Ar(\nu,\nu+p)$ cross section as a function of the kinetic
  energy of the final proton. The dashed histogram shows results
  without rescattering (PWIA) and the solid one has been obtained from
  a MC cascade simulation.  }\label{fig:rpa-fsi}
\end{figure}

The double differential cross section, with respect to the outgoing
lepton kinematical variables,  for the process $\nu_l (k) +\, A_Z \to
l^- (k^\prime) + X $  is given in the Laboratory (LAB) frame
by\footnote{Extensions to antineutrino or NC induced processes are
  straightforward. Details can be found in Refs.~\cite{ccjuan,ncjuan}.}
\begin{equation}
\frac{d^2\sigma_{\nu l}}{d\Omega(\hat{k^\prime})dE^\prime_l} =
\frac{|\vec{k}^\prime|}{|\vec{k}~|}\frac{G^2}{4\pi^2} 
L_{\mu\sigma}W^{\mu\sigma} \label{eq:sec}
\end{equation}
with $\vec{k}$ and $\vec{k}^\prime~$ the LAB lepton momenta, $G$ the
Fermi constant and $L$ and $W$ the leptonic and hadronic tensors,
respectively.  The hadronic tensor includes all sort of non-leptonic
vertices and is determined by the $W^+-$boson selfenergy,
$\Pi^{\mu\rho}_W(q)$, in the nuclear medium. We follow here the
formalism of Ref.~\cite{GNO97}, and we evaluate the selfenergy of a
neutrino moving in infinite nuclear matter of density $\rho$. We
obtain,
\begin{eqnarray}
W^{\mu\sigma}_s (q) &\propto&  \Theta(q^0) 
\int \frac{d^3 r}{2\pi}~ {\rm Im}\left [ \Pi_W^{\mu\sigma} 
+ \Pi_W^{\sigma\mu} \right ] (q;\rho(r))\label{eq:wmunus}\\
W^{\mu\sigma}_a (q) &\propto&   \Theta(q^0) 
 \int \frac{d^3 r}{2\pi}~{\rm Re}\left [ \Pi_W^{\mu\sigma} 
- \Pi_W^{\sigma\mu}\right] (q;\rho(r)) \label{eq:wmunua}
\end{eqnarray}
with $W^{\mu\sigma}= W^{\mu\sigma}_s + {\rm i} W^{\mu\sigma}_a$,
$q=k-k'$, and where we have used the Local Density Approximation
(LDA), which assumes a FG model for the nucleus to start
with\footnote{ Large basis shell model schemes provides a very
  accurate description of the nuclear ground state wave
  functions~\cite{SM}, which is unnecessary when one is dealing with
  inclusive processes and nuclear excitation energies above, let us
  say, 50 MeV~\cite{capture}. Besides, the description of high-lying
  excitations necessitates the use of large model spaces and this
  often leads to computational difficulties, making the approach
  applicable essentially only for neutrino energies in the range of
  tens of MeV.}. The virtual $W$ gauge boson can be absorbed by one
nucleon, 1p1h nuclear excitation, leading to the QE contribution to
the nuclear response function. In this case, the $W-$selfenergy is
determined, besides the $W^\pm NN$ vertex, by the imaginary part of
isospin asymmetric Lindhard function. We work on an non-symmetric
nuclear matter with different Fermi sea levels for protons 
than for neutrons.  Explicit expressions can be found in
~\cite{ccjuan}. In what follows, we will consider further improvements
on this simple framework:
\begin{itemize}
\item We enforce a \underline{correct energy balance} of the
  different studied processes and consider the effect of the
  \underline{Coulomb} field of the nucleus acting on the ejected
  charged lepton.

\item \underline{RPA and SRC}: We take into account polarization
  effects by substituting the particle-hole (1p1h) response by an RPA
  response consisting of a series of ph and $\Delta$h excitations.  We
  use a Landau-Migdal ph-ph interaction~\cite{Sp77}: $V =
  c_{0}\left\{ f_{0}+f_{0}^{\prime}\vec{\tau}_{1}\vec{\tau}_{2}+
    g_{0}\vec{\sigma}_{1}\vec{\sigma}_{2}+g_{0}^{\prime}
    \vec{\sigma}_{1}\vec{\sigma}_{2} \vec{\tau}_{1}\vec{\tau}_{2}
  \right\}$.  In the vector-isovector channel ($\vec{\sigma}
  \vec{\sigma} \vec{\tau} \vec{\tau}$ operator) we use an
  interaction~\cite{GNO97} with explicit $\pi-$meson (longitudinal)
  and $\rho-$meson (transverse) exchanges, that also includes SRC and
  $\Delta(1232)$ degrees of freedom.  RPA effects are extremely
  important, as confirmed by several groups~\cite{crpa}, and should be
  definitely taken into account in any neutrino oscillation
  analysis~\cite{nuint07}. As a matter of example, we show in the left
  panel of Fig.~\ref{fig:rpa-fsi} results  in
  $^{16}$O at intermediate energies~\cite{ccjuan}.

\item \underline{SF+FSI:} We take into account the modification of the
  nucleon dispersion relation in the medium by using nucleon
  propagators properly dressed with a realistic
  self-energy~\cite{FO92}.  Thus, we compute the imaginary part of the
  Lindhard function (ph propagator) using realistic particle and hole
  SF's.  The effect is twofold, firstly by using the hole SF, we go
  beyond a simple FG of non-interacting nucleons, and we include some
  interactions among the nucleons. Secondly, the particle SF accounts
  for the interaction of the ejected nucleon with the final nuclear
  state; this is most commonly called Final State Interaction (FSI) in
  the literature. We show some results in the left panel of
  Fig.~\ref{fig:rpa-fsi}, taken from Ref.~\cite{ccjuan}. We find a
  sizeable reduction of the strength at the QE peak, which is slightly
  shifted, and an enhancement of the high energy transfer tail. For
  integrated cross sections both effects partially compensate.  We
  find a qualitative and quantitative agreement with the results of
  Benhar et al.~\cite{sakuda} and of the Giessen
  group~\cite{Buss:2007ar}.
\end{itemize}
We have estimated the theoretical uncertainties of our model by Monte
Carlo (MC) propagating the uncertainties of its different inputs into
differential and total cross sections~\cite{ccjuan-errors}.  We
conclude that our approach provides QE $\nu(\bar \nu)$--nucleus cross
sections with relative errors of about
10-15\%, while uncertainties affecting the ratios
$\sigma(\mu)/\sigma(e)$ and $\sigma(\bar\mu)/\sigma(\bar e)$ would be
certainly smaller, not larger than about 5\%, and mostly coming
from deficiencies of the local FG picture of the
nucleus~\cite{ccjuan-errors}. 

Finally in the QE region, we have also studied CC and NC nucleon
emission processes which play an important role in the analysis of
oscillation experiments. In particular, they constitute the unique
signal for NC neutrino driven reactions. We use a MC simulation method
to account for the rescattering of the outgoing nucleon~\cite{GNO97}.
The first step is the gauge boson ($W^{\pm}$ and $Z^0$ ) absorption in
the nucleus\footnote{Some calculations in the literature use the PWIA
  and DWIA, including or not relativistic effects.  The PWIA
  constitutes a poor approximation, since it neglects all types of
  interactions between the ejected nucleon and the residual nuclear
  system. The DWIA describes the ejected nucleon as a solution of the
  Dirac or Schr\"odinger equation with an optical potential obtained
  by fitting elastic proton--nucleus scattering data. The imaginary
  part accounts for the absorption into unobserved channels. This
  scheme is incorrect to study nucleon emission processes where the
  state of the final nucleus is totally unobserved, and thus all final
  nuclear configurations, either in the discrete or on the continuum,
  contribute.  The distortion of the nucleon wave function by a
  complex optical potential removes all events where the nucleons
  collide inelastically with other nucleons.  Thus, in DWIA
  calculations, the nucleons that interact inelastically are lost when
  in the physical process they simply come off the nucleus with a
  different energy, angle, and maybe charge, and they should
  definitely be taken into account.}. Different distributions for both
NC and CC processes can be found in ~\cite{ncjuan}, as example, we
show here results for NC nucleon emission from argon (right panel of
Fig.~\ref{fig:rpa-fsi}). The rescattering of the outgoing nucleon
produces a depletion of the high energy side of the spectrum, but the
scattered nucleons clearly enhance the low energy region. Our results
compare well with those of Ref.~\cite{Buss:2007ar} obtained by means
of a transport model.
\begin{figure}[th]
\includegraphics[width=6.7cm,height=4.5cm]{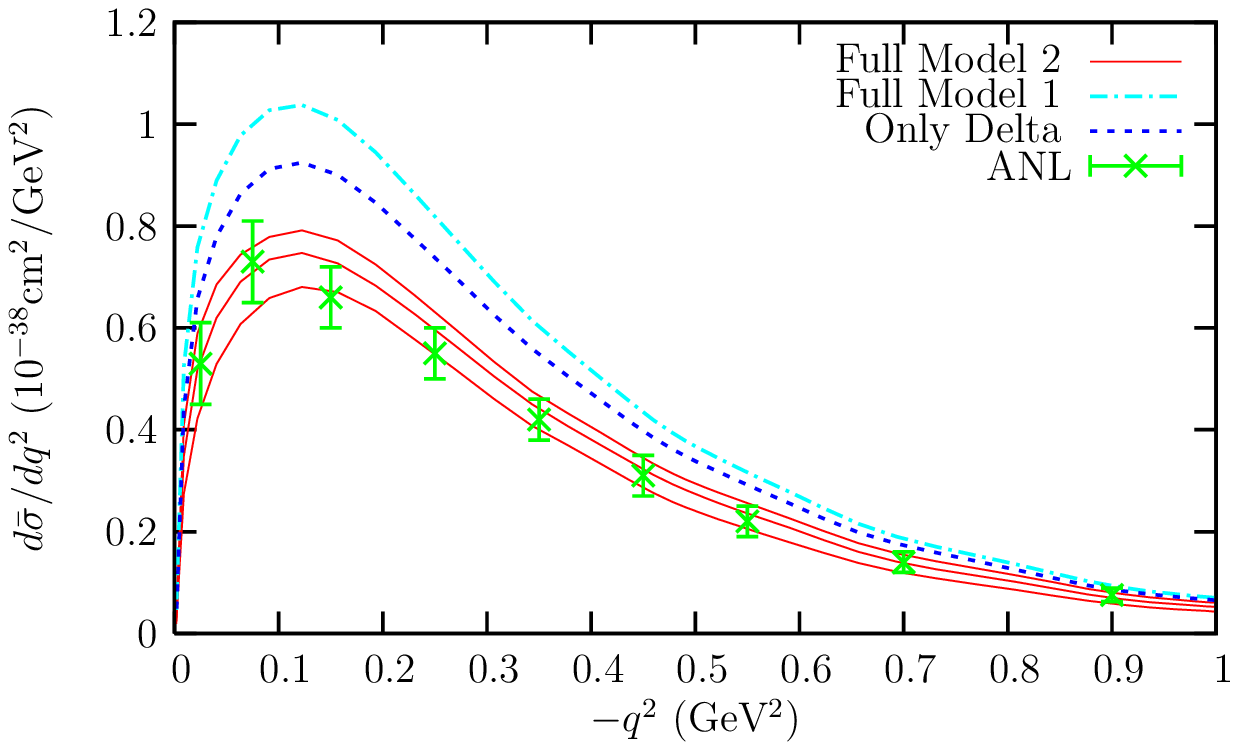}\hspace{1cm}
\includegraphics[width=6.7cm,height=4.5cm]{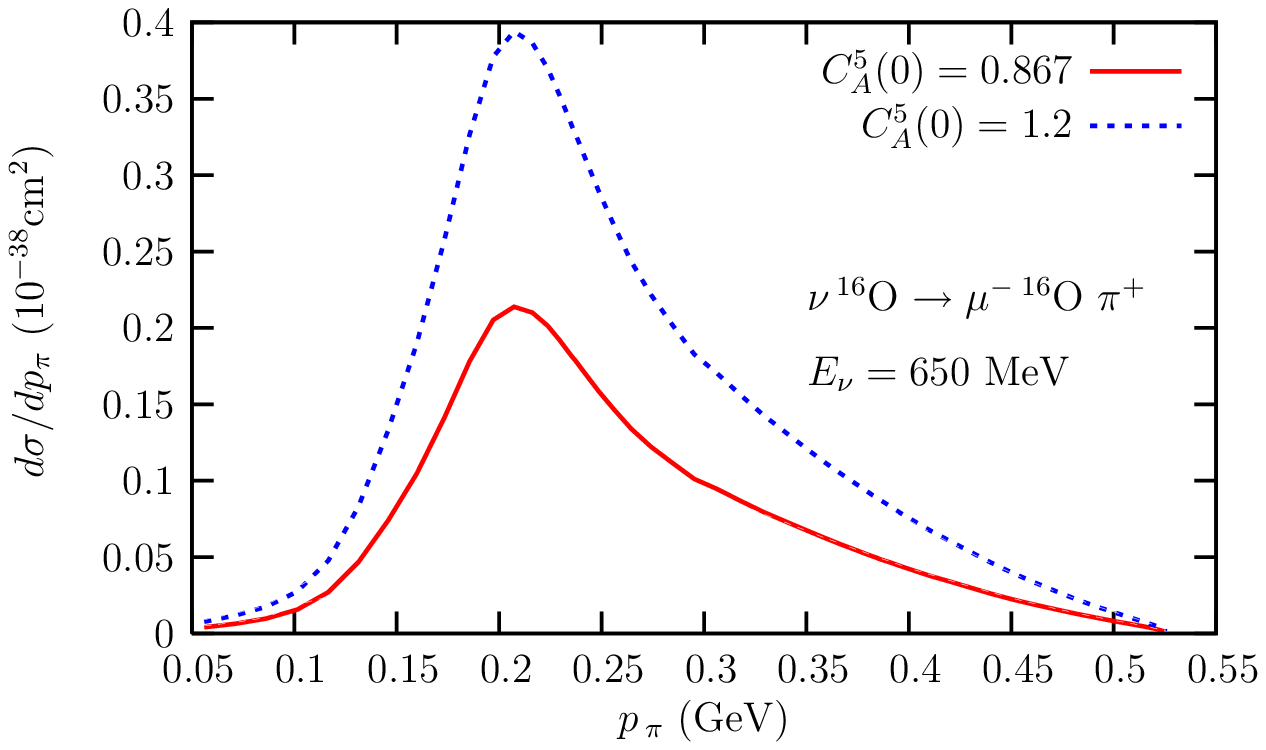}
\caption{ Left: Flux averaged $\pi N$ invariant mass distribution of events
  for the $\nu_\mu p \to \mu^- p \pi^+$ reaction.  Dashed lines stand
  for the contribution of the $\Delta$ pole term with $C_5^A(0)=1.2$
  (GTR) and $M_{A\Delta}= 1.05$ GeV.  Dashed--dotted and central solid
  lines are obtained when the full model of Ref.~\cite{prd} is
  considered with $C_5^A(0)=1.2,\, M_{A\Delta}= 1.05$ GeV
  (dashed-dotted) and with the best fit parameters $C_5^A(0)=0.867,\,
  M_{A\Delta}= 0.985$ GeV (solid). For this latter case, we also show
  the 68\% CL bands. 
  Right:  CC  coherent pion production differential cross section.
 }\label{fig:res5}
\end{figure}

\section{Chiral Symmetry and Neutrino Pion 
Production off the Nucleon}
The neutrino pion production off the nucleon is traditionally
described in the literature by means of the weak excitation of the
$\Delta(1232)$ resonance and its subsequent decay into $N\pi$.  Here,
we present results from a model~\cite{prd} that includes also some
background terms required by chiral symmetry.  The contribution of
these terms is sizeable and leads to significant effects in total and
partially integrated pion production cross sections at intermediate
energies. We re-adjust the $C_5^A(q^2)$ form--factor, that controls
the largest term of the $\Delta-$axial contribution, and find
corrections of the order of 30\% to the off diagonal
Goldberger-Treiman relation (GTR), when the $\nu_\mu p \to \mu^-p\pi^+$ ANL
 $q^2-$differential cross section data~\cite{anl} are
fitted (right panel of Fig.~\ref{fig:res5}). Thus,  we find a
substantially smaller contribution of the $\Delta$ pole mechanism than
in other approaches~\cite{weak-pi}, which has an important effect on the 
CC and NC nuclear coherent pion production cross sections 
(Fig.~\ref{fig:res5}). We have also extended the model to describe two pion
production processes near threshold~\cite{twopion}.

\end{document}